\documentclass[pre,twocolumn,nofootinbib,twoside,showpacs,superscriptaddress,tightenlines,floatfix]{revtex4-2}
\usepackage{dcolumn}
\usepackage{lipsum}
\makeatletter
\let\old@makecaption=\@makecaption
\usepackage{subcaption}
\let\@makecaption=\old@makecaption
\makeatother
\usepackage{booktabs}
\usepackage{graphicx,tabularx,amssymb,amsfonts,amsmath,epsf,bm}
\usepackage[utf8x]{inputenc}
\usepackage[T1]{fontenc}
\usepackage[normalem]{ulem}
\usepackage{color}
\usepackage{mathtools}
\usepackage{dsfont}
\usepackage{nicefrac}

\usepackage{cancel}
\usepackage{hyperref}
\hypersetup{
    allcolors=blue,
    colorlinks=true,
    bookmarks=true,
    pdfpagemode=FullScreen,
}
\definecolor{nred}{RGB}{224,0,0}
\definecolor{nblue}  {RGB}{28,130,185}
\definecolor{dgreen} {RGB}{38,150,21}
\definecolor{norange}{RGB}{230,120,20}

\begin{document}

\title{Convergence of the conformal Ward identity in the derivative expansion approximation
}

\author{Jorge Ibañez}
\affiliation{Instituto de F\'isica, Facultad de Ingenier\'ia, Universidad de la Rep\'ublica, J.H.y Reissig 565, 11000 Montevideo, Uruguay}%

\author{Matthieu Tissier}
\affiliation{Sorbonne Universit\'e, CNRS, Laboratoire de Physique Th\'eorique de la Mati\`ere Condens\'ee, LPTMC, 75005 Paris, France}%

\author{Gonzalo De Polsi}
\affiliation{Instituto de F\'isica, Facultad de Ciencias, Universidad de la Rep\'ublica, Igu\'a 4225, 11400, Montevideo, Uruguay}%

\date{\today}

\begin{abstract}
Conformal invariance is expected to be an emergent property of many systems in their critical regime. However, approximation schemes generically spoil this property. This is in particular the case of the derivative expansion, a widely used approximation scheme in the framework of the functional renormalization group.  In this article, we consider Ward identities associated with conformal invariance in the 3-d Ising universality class with truncations at order 4 (next-to-next-to-leading order) in the derivative expansion, with $Z_2$ invariant composite operators. Our results confirm that the regulating functions which yield a small breaking of conformal invariance also present a small sensitivity of the universal critical exponents with the choice of this regulating function. We also show that, in the vicinity of regulator-parameter values for which the conformal constraints are best satisfied, the breaking of conformal invariance reduces as the order of the derivative expansion is increased, providing a new indication of the convergence of this approximation scheme.
\end{abstract}

\maketitle

\section{Introduction}

Continuous phase transitions constitute one of the most remarkable manifestations of universality in many-body systems. Near criticality, microscopic details become irrelevant, and long-distance physics is governed by scale invariance, leading to a universal behavior characterized by critical exponents and scaling functions. Already in the seminal works of Wilson and Kadanoff \cite{WILSON197475,PhysRevB.4.3174,PhysicsPhysiqueFizika.2.263}, it became clear that renormalization group (RG) ideas provide the natural framework to explain universality and scaling. A striking consequence of this paradigm is that, at criticality, scale invariance is often enhanced to full conformal invariance. Conformal symmetry severely constrains correlation functions and operator spectra, thereby sharpening our understanding of critical phenomena.

Historically, conformal symmetry first played a decisive role in two-dimensional systems. The infinite-dimensional conformal algebra in two dimensions led to the exact solution of numerous critical models, culminating in the development of conformal field theory (CFT) by Belavin, Polyakov, and Zamolodchikov \cite{BPZ1984}, and its systematic formulation, see \cite{DiFrancesco1997}. In higher dimensions, conformal symmetry is finite-dimensional, yet it still imposes powerful constraints. Early work by Polyakov \cite{Polyakov1970} and later developments \cite{Polchinski1988,OsbornPetkou1994} clarified the structure of conformal correlation functions and the operator product expansion in general dimensions. The question of whether scale invariance implies conformal invariance has been extensively investigated \cite{Polchinski1988,Nakayama2015,Delamotte2016, Cabrera2026}, with strong evidence supporting equivalence under broad conditions.

In the context of statistical mechanics, paradigmatic universality classes such as the three-dimensional Ising model are believed to be described by nontrivial conformal field theories at criticality. In the past two decades, the conformal bootstrap program has revolutionized the quantitative determination of critical exponents and operator dimensions in this universality class \cite{ElShowk2012,Kos2014,SimmonsDuffin2015}. In particular, high-precision estimates for the 3D Ising universality class have been obtained using numerical bootstrap techniques \cite{ElShowk2012,Simmons2017,Chang2025}. These results are in excellent agreement with Monte Carlo simulations \cite{Hasenbusch2010,Ballesteros1996,Engels1997,Ferrenberg2018,Kaupuzs2023} and high-order perturbative computations in the $\epsilon$-expansion and fixed-dimension approaches \cite{WilsonFisher1972,Baker1978,ZinnJustin2001,Calabrese2003,PelissettoVicari2002,Kompaniets2017}.

From the field-theoretic perspective, the $\phi^4$ theory provides the canonical description of the Ising universality class near four dimensions. Perturbative RG analyses around $d=4$ via the $\epsilon$-expansion \cite{WilsonFisher1972} and high-order resummations in fixed dimension \cite{ZinnJustin2001,Calabrese2003,Kompaniets2017} have yielded accurate predictions for critical exponents. However, perturbation theory is intrinsically tied to small couplings or dimensional expansions. Nonperturbative approaches are therefore essential to address strongly coupled fixed points directly in three dimensions.

Among these, the functional renormalization group (FRG) has emerged as a versatile and powerful framework. The modern formulation stems from Wilson’s insight of integrating out fluctuations in successive momentum shell \cite{WILSON197475}, and was cast into an exact functional differential equation by Polchinski \cite{Polchinski1984}. An alternative but equivalent formulation, particularly suited for nonperturbative approximations, is the Wetterich equation for the scale-dependent effective average action \cite{Wetterich1993}. Reviews of the FRG formalism and its applications can be found in \cite{Delamotte2012,Berges2002,Pawlowski2007,Rosten2012,Dupuis2021}.

The central object in the FRG approach is the effective average action $\Gamma_k$, which interpolates between the microscopic action in the ultraviolet and the full effective action in the infrared. Its RG flow is governed by an exact one-loop equation with a regulator function implementing Wilsonian coarse-graining \cite{Wetterich1993}. In practice, solving this functional equation requires truncations. One of the most widely used and systematically improvable approximation schemes is the derivative expansion (DE) \cite{Morris1994}. In this scheme, the effective average action is expanded in powers of derivatives of the fields, while retaining a functional dependence on the field itself. A truncation retaining all terms with up to \(n\) derivatives is denoted by \(O(\partial^n)\). At leading order, known as the local potential approximation (LPA), only a scale-dependent effective potential and an unrenormalized kinetic term are kept. At next-to-leading order (NLO), corresponding to \(O(\partial^2)\), a field-dependent wavefunction renormalization is included, and at next-to-next-to-leading order (NNLO), $O(\partial^4)$ three more functions are incorporated \cite{Morris2005}, see below for more detail.

The derivative expansion has proven remarkably successful in computing critical exponents and scaling functions in a wide range of systems \cite{Litim2001,Litim2002,Canet2003,DePolsi2020a,DePolsi2021a,Chlebicki2022,DePolsi2026}. For the three-dimensional Ising universality class, successive orders in the derivative expansion lead to results in good agreement with Monte Carlo simulations and conformal bootstrap estimates. 
However, to do so, this approximation scheme is supplemented with the principle of minimal sensitivity (PMS) \cite{Canet2003,Canet2005}. The PMS relates to the fact that approximate results within the FRG exhibit a dependence on the choice of regulator. Although physical observables are regulator-independent in the exact theory, truncations introduce residual scheme dependence which must be fixed in some way. The PMS relies on the expected behavior of the exact theory to fix the regulating function to the case that yields the least dependence upon small variations of the function. Once the PMS is implemented, the DE
significantly improves its convergence properties and numerical stability \cite{Balog2019,DePolsi2020a,DePolsi2022}. 

Despite its successes, the derivative expansion is not manifestly compatible with all symmetries of the underlying theory. In particular, while scale invariance at fixed points is naturally implemented in the dimensionless formulation of the flow, full conformal invariance is not guaranteed within truncated flows. Conformal symmetry implies additional constraints beyond scale invariance, such as the vanishing of the virial current and the realization of special conformal Ward identities \cite{Polchinski1988,OsbornPetkou1994,Delamotte2016}. The compatibility between FRG approximations and conformal symmetry has therefore become an important conceptual and practical issue.

In recent years, the realization of conformal symmetry within the FRG framework has been investigated in detail. In this scenario, Ward identities may be modified due to the presence of the regulator. Modified Ward identities associated with scale and conformal transformations have been derived in the presence of the regulator \cite{Sonoda1993,Rosten2012}, and the conditions for conformal invariance at critical fixed points have been analyzed both formally and within concrete truncations \cite{Balog2020,Delamotte2021}. In particular, the interplay between regulator dependence, scale invariance and the emergence of full conformal symmetry has been clarified in a series of works \cite{Balog2020,Delamotte2021,Cabrera2025}, highlighting how truncations may induce spurious symmetry breaking effects.

Within the derivative expansion, violations of conformal constraints have been explicitly analyzed at leading nontrivial order \cite{Balog2019,Delamotte2021,Cabrera2025}. These studies revealed that while scale invariance is automatically realized at fixed points, special conformal constraints are generally violated at finite order of the DE.

The PMS \cite{Stevenson1981,Canet2005} and related optimization criteria \cite{Litim2001,Litim2002} provide systematic ways to select regulator functions that minimize the dependence of approximate results on the regulating function. Interestingly, it has been observed that regulator choices favored by the PMS also tend to improve the fulfillment of conformal Ward identities \cite{Balog2020,Delamotte2021,Cabrera2025}, suggesting a nontrivial connection between optimization and realization of the conformal symmetry. 
This observation led to the principle of maximal conformality (PMC) as another possible criterion to select the regulator function by demanding it to be chosen as the one that best realizes conformal symmetry constraints.

In this work, we revisit the realization of conformal symmetry within the derivative expansion of the FRG applied to the $\phi^4$ theory. Extending previous analyses performed at NLO \cite{Delamotte2021} and at NNLO but without the inclusion of composite operators \cite{Balog2020}, we consider the derivative expansion at NNLO, $O(\partial^4)$ with the addition of a source for composite operators. At this order, new conformal constraints become accessible for study, and it becomes possible to track the same constraint studied in \cite{Delamotte2021} across consecutive orders of the approximation scheme. This provides a stringent test of the convergence of the derivative expansion with respect to symmetry restoration. Our results show that conformal constraints are increasingly well satisfied at higher orders, suggesting a progressive restoration of conformal symmetry as higher-derivative operators are included (see, however, \cite{Morris2026} for a discussion of convergence at very large orders). We also show that conformal constraints are optimally fulfilled for regulator choices consistent with the principle of minimal sensitivity, thereby further clarifying the interplay between regulator optimization and emergent conformal symmetry at criticality.

This paper is organized as follows. In Sec.~\ref{Sec:FRG} we recall the basics of FRG and present the ideas and ansatz of the DE used within this work. We then discuss how conformal symmetry is treated within this framework in Sec.~\ref{Sec:Symm}. The results and comparison to previous works are presented in Sec.~\ref{Sec:Res} and we give our conclusions in Sec.~\ref{Sec:Concl}.

\section{Functional Renormalization Group and the Derivative Expansion}\label{Sec:FRG}

The functional renormalization group provides a non-perturbative realization of Wilson's renormalization group in terms of a scale-dependent effective action, the \emph{effective average action} $\Gamma_k[\varphi,K]$.  For the purpose of this work, we include a dependence on a source $K$ which couples to scaling operators which are scalar, $Z_2$-invariant composite operators $\mathcal{O}$ through $\int_x K(x)\mathcal{O}(x)$. In this context, $\Gamma_k[\varphi,K]$ interpolates smoothly between the microscopic (bare) action $S$ at a UV cutoff scale $k=\Lambda$ and the full effective action $\Gamma[\varphi,K]$ as $k\to 0$. Infrared modes with momenta $p\ll k$ are suppressed by the introduction of a regulator term $R_k(p^2)$, which acts as a momentum-dependent mass, while modes with $p\gtrsim k$ remain essentially unaffected. The evolution of $\Gamma_k$ with the RG scale is described exactly by the Wetterich equation~\cite{Wetterich1993}:

\begin{equation}\label{eq:Wetterich}
\partial_t \Gamma_k[\varphi,K] = \frac{1}{2} \mathrm{Tr} \big[ \big(\Gamma_k^{(2,0)}[\varphi,K] + R_k\big)^{-1} \partial_t R_k \big],
\end{equation}
where the trace stands for an integral over volume, and eventually internal indices summation, $\Gamma_k^{(2,0)}$ is the second functional derivative with respect to the field $\varphi$  with the inverse understood in the operator sense and $t=\ln(k/\Lambda)$ is the RG ``time''.  Despite its one-loop structure, Eq.~\eqref{eq:Wetterich} is exact and incorporates the flow of all operators compatible with the symmetries of the theory. We denote by $X^{(n,m)}$ the functional derivative of $X$ with respect to the field $\varphi$ ($n$ times) and with respect to the source $K$ ($m$ times).

The regulator function $R_k(p^2)$ plays a central role in the FRG. To satisfy the requirements of the effective average action framework, it must provide a mass-like cutoff for low-momentum modes $p^2\ll k^2$ to suppress infrared divergences and vanish sufficiently rapidly for $p^2\gg k^2$ so that high-momentum modes are integrated out without distortion.

In order to implement the RG ideas to study the critical theory, one must rescale the theory to look for fixed point solutions to the dimensionless flow equation. This is achieved by implementing the following change of variables to measure quantities in terms of the scale $k$:
\begin{equation}\label{eq:rescaling}
    \begin{aligned}
    	x &= k^{-1}\tilde{x}, \\
    	\varphi(x) &= k^{(d-2)/2} Z_k^{-1/2}\,\tilde{\varphi}(\tilde{x}), \\
    	K(x) &= k^{d-D_{\mathcal{O}}}\,\tilde{K}(\tilde{x}).
    \end{aligned}
\end{equation}
With this change, the dimensionless flow equation then becomes:

\begin{equation}\label{eq:WettEqDless}
	\begin{split}
		\partial_t \Gamma_k[\tilde{\varphi},\tilde{K}]=&\int_{\tilde{x}}\frac{\delta \Gamma_k}{\delta \tilde{\varphi}(\tilde{x})}\big(\tilde{x}^\nu\tilde{\partial}_\nu+\frac{d-2+\eta_k}{2}\big)\tilde\varphi(\tilde{x})\\&-\int_{\tilde{x}} \tilde{K}(\tilde{x})\big( \tilde{x}^\nu \tilde{\partial}_\nu + D_{\mathcal O} \big)\frac{\delta \Gamma_k}{\delta \tilde{K}(\tilde{x})}\\&+\frac{1}{2}\int_{\tilde{x},\tilde{y}}\tilde{\partial_t R_k}(\tilde{x},\tilde{y})\tilde{G}^{(2,0)}_k(\tilde{x},\tilde{y}),
    \end{split}
\end{equation}
where $\tilde{\partial_t R_k}$ represents the dimensionless version of the function $\partial_tR_k$ in direct space, and $D_{\mathcal O}$ denotes the scaling dimension of the composite operator $\mathcal O$.
The dimensionless propagator $\tilde{G}^{(2,0)}_k$ stands for the dimensionless version of $\big(\Gamma_k^{(2,0)}[\varphi,K] + R_k\big)^{-1}$. At its critical point, the physical theory flows toward a fixed point for which $\eta_k^*=\eta$ becomes the anomalous dimension of the field that relates to the scaling dimension of the field in the familiar way $D_\varphi=(d-2+\eta)/2$. When looking for fixed points, one sets:
\begin{equation}\label{eq:FixPoint}
	\partial_t \Gamma_k[\tilde{\varphi},\tilde{K}]=0.
\end{equation}
Once the fixed point is obtained, critical exponents are determined by linearizing the RG flow around it. In this work, we focus on the perturbations associated with the critical exponents $\nu$ and $\omega$.

The Wetterich equation, being exact, implies that physical quantities such as critical exponents or critical temperature should be independent of the regulating function $R_k$. This is true at the level of the exact equations ~\eqref{eq:Wetterich} or \eqref{eq:WettEqDless}, but fails to be true when unavoidable\footnote{Exact solutions of Eq.~\eqref{eq:Wetterich} or Eq.~\eqref{eq:WettEqDless} are infeasible in practice due to its functional and non-linear nature.} approximations are performed.
To minimize these truncation artifacts and improve the efficiency of numerical integrations, optimized regulators are often employed supported by some criterion, such as PMS or PMC.

The Fourier transform of the regulator is, generically, written as:
\begin{equation}\label{eq:regProfile}
    R_k(q^2)=\alpha Z_k k^2 r(q^2/k^2).
\end{equation}
Typical choices include the exponential regulator given in 
Eq.~\eqref{eq:regulator-exp}, the Wetterich regulator Eq.~\eqref{eq:regulator-wetterich} and the Litim-optimized regulator~\cite{Litim2001} or its generalizations \cite{Balog2019}, Eq.~\eqref{eq:regulator-theta}.

\begin{align}
E_{k}(q^2) &= \alpha Z_k k^2 \,\exp(-q^2/k^2) , \label{eq:regulator-exp}\\
W_k(q^2) &= \alpha Z_k k^2 \frac{q^2/k^2}{\exp(q^2/k^2) - 1}, \label{eq:regulator-wetterich}\\ 
\Theta^n_k(q^2) &= \alpha Z_k k^2 \, (1-q^2/k^2)^n \theta(1-q^2/k^2).  
\label{eq:regulator-theta}
\end{align}

In what follows we discuss one of the most used approximation schemes within this framework, the \emph{derivative expansion}, where the effective average action is expanded in powers of derivatives of the field while retaining full field dependence in the expansion coefficients. For a single scalar field, the derivative expansion at order $O(\partial^4)$ in the presence of a source for composite operators which appears at most linearly can be written as
\begin{widetext}
\begin{equation}\label{Eq:ansatz}
    \begin{split}
\Gamma_{k}[\varphi, K]= & \int_{x} \Big\lbrace U_{0}(\varphi)+ K(x)U_{1}(\varphi)+\frac{1}{2}\big[Z_{0}(\varphi)+K(x) Z_{1}(\varphi)\big](\partial_{\mu} \varphi)^{2}+\frac{1}{2}\big[W_{a 0}(\varphi)+K(x) W_{a 1}(\varphi)\big](\partial_{\mu} \partial_{\nu} \varphi)^{2} \\
& +\frac{1}{2}\big[W_{b 0}(\varphi)+K(x) W_{b 1}(\varphi)\big] \varphi(\partial_{\mu} \varphi)(\partial_{\nu} \varphi)(\partial_{\mu} \partial_{\nu} \varphi) +\frac{1}{2}\big[W_{c 0}(\varphi)+K(x) W_{c 1}(\varphi)\big](\partial_{\mu} \varphi)^{2}(\partial_{\nu} \varphi)^{2} \\
& +Y_{a}(\varphi) \varphi(\partial_{\mu} K(x))(\partial_{\mu} \varphi)+Y_{b}(\varphi) \varphi(\partial_{\mu} \partial_{\nu} K(x))(\partial_{\mu} \partial_{\nu} \varphi) +Y_{c}(\varphi)(\partial_{\mu} K(x))(\partial_{\nu} \varphi)(\partial_{\mu} \partial_{\nu} \varphi)\\
& +Y_{d}(\varphi)(\partial_{\mu} \varphi)(\partial_{\nu} \varphi)(\partial_{\mu} \partial_{\nu} K(x))+Y_{e}(\varphi) \varphi(\partial_{\mu} \varphi)^{2}(\partial_{\nu} \varphi)(\partial_{\nu} K(x))\Big\rbrace ,
    \end{split}
\end{equation}
\end{widetext}
where we omit the dependence of functions on the scale $k$, $U_0(\varphi)$ is the running effective potential, $Z_0(\varphi)$ encodes the leading momentum dependence of the two-point function of the field, and so on. Notice that at $K=0$, one recovers the typical ansatz used in previous implementations of the DE \cite{Canet2005,Balog2019,Balog2020} for this universality class. The terms proportional to $K$ include all terms allowed by the internal symmetries.

At the lowest order, the \emph{local potential approximation}, one sets $Z_k \equiv 1$ and neglects higher derivatives, reducing Eq.~\eqref{eq:Wetterich} to a single partial differential equation for $U_k(\varphi)$. Despite its simplicity, LPA captures the essential fixed-point structure and allows qualitative estimates of critical exponents while setting the critical exponent $\eta$ associated with field anomalous dimension to zero. Including $Z_k(\varphi)$ at order $O(\partial^2)$ permits a nonzero anomalous dimension $\eta$ and significantly improves quantitative predictions. Higher orders of the derivative expansion incorporate terms with additional derivatives. The highest order reported in the literature involves six-derivative terms, with \cite{Balog2019} providing the sole implementation at this level. Systematic inclusion of these terms allows controlled improvement of predictions and assessment of truncation errors.

Substituting the derivative expansion ansatz~\eqref{Eq:ansatz} into the Wetterich equation and projecting onto the various derivative structures yields a closed set of flow equations for the running functions $U_0(\varphi)$, $Z_0(\varphi)$, and so on. The projection is typically done by evaluating the flow at constant field (for $U_0$) or by extracting coefficients in a momentum expansion (for $Z_0$ extracting the $p^2$ structure of the two-point function). Numerical integration of these coupled equations from the UV scale $k=\Lambda$ down to the IR scale $k\to 0$ allows the determination of fixed-point potentials, anomalous dimensions, and universal critical exponents.

The convergence of the derivative expansion has been studied extensively. Theoretical arguments and numerical tests indicate that, for smooth and appropriately chosen regulators, the expansion converges rapidly for a wide range of models, particularly near nontrivial fixed points such as the Wilson-Fisher fixed point \cite{Balog2019,DePolsi2020a,Peli2021,DePolsi2021a,Chlebicki2022,DePolsi2022}.  

Empirically, the LPA already provides qualitatively correct fixed-point structure. The inclusion of $O(\partial^2)$ terms improves critical exponents typically to a few percent accuracy, and further orders reduce discrepancies with Monte Carlo simulations, perturbative results and other field theoretical approaches. 
The derivative expansion is thus a systematically improvable approximation: each additional order captures new momentum-dependent interactions and refines the estimates of universal quantities typically by a factor of 4--9 \cite{Balog2019}.  

Nevertheless, one must keep in mind that some aspects are still unresolved. Although empirically the DE yields more accurate and precise results as one goes to higher orders, there is no formal proof for its convergence other than educated and reasonable argumentation. Additionally, the \emph{principle of minimal sensitivity} \cite{Canet2005,Balog2019,DePolsi2022}, which consists of varying the regulator profile and picking quantities at values where less dependence is exhibited on the regulator, plays a crucial role in good behavior to kick-in. Indeed, this is the major drawback of pushing the DE to higher orders, since it is evidenced that, even though the precision increases with increasing order, the actual dependence on the regulator profile also does. This implies that correctly selecting the regulator profile becomes increasingly consequential.

In summary, the Wetterich FRG framework combined with the derivative expansion provides a controlled, systematically improvable, and non-perturbative approach to studying renormalization group flows. With a careful choice of regulator and truncation scheme, it allows for the extraction of accurate universal quantities and offers insights into the structure of field theories beyond the reach of traditional perturbative techniques.

In this work we will study the realization of the expected conformal symmetry with the regulator profile and how this relates to the PMS criterion. To this end, in the next section we address how conformal symmetry transformations are included in this formalism.

\section{Conformal Transformations and Ward Identities within the FRG}\label{Sec:Symm}

In this section, we describe how the symmetries of the conformal group  are treated within the FRG framework, for a more detailed discussion see \cite{Delamotte2021,Cabrera2025,DePolsiThesis}. We will assume the action and the integration measure in the functional integral to be invariant under all conformal transformations: translations, rotations, dilatations, and special conformal transformations (SCT). We then obtain the modified Ward identities. The regulating term of FRG breaks conformal and dilatation symmetry, but leaves the translation and rotation Ward identities untouched.

The corresponding infinitesimal transformations are realized by the following variations of the field and the scalar composite operator:
\begin{equation}\label{eq:TransfField}
	\begin{split}
		\delta_{\rm tra}\,\phi(x)&=\epsilon_\mu \partial_\mu\phi(x),\\
        \delta_{\rm tra}\,\mathcal{O}(x)&=\epsilon_\mu \partial_\mu\mathcal{O}(x),\\
		\delta_{\rm rot}\,\phi(x)&=\epsilon_{\mu\nu} \big(x_\mu \partial_\nu-x_\nu \partial_\mu\big) \phi(x),\\
		\delta_{\rm rot}\,\mathcal{O}(x)&=\epsilon_{\mu\nu} \big(x_\mu \partial_\nu-x_\nu \partial_\mu\big) \mathcal{O}(x),\\
		\delta_{\rm dil}\,\phi(x)&=\epsilon \big(x_\mu \partial_\mu+D_\varphi\big) \phi(x),\\
		\delta_{\rm dil}\,\mathcal{O}(x)&=\epsilon \big(x_\mu \partial_\mu+D_\mathcal{O}\big) \mathcal{O}(x),\\
		\delta_{\rm conf}\, \phi(x)&=\epsilon_\mu \big(x^2\partial_\mu-2x_\mu x_\nu\partial_\nu-2x_\mu D_\varphi\big) \phi(x),\\
		\delta_{\rm conf}\, \mathcal{O}(x)&=\epsilon_\mu \big(x^2\partial_\mu-2x_\mu x_\nu\partial_\nu-2x_\mu D_\mathcal{O}\big) \mathcal{O}(x),\\
	\end{split}
\end{equation}
where $D_{\mathcal O}$ is the dimension of the primary operator $\mathcal O$.

From this point onward, following the procedure given in \cite{Delamotte2021,Cabrera2025}, it is straightforward to obtain the expressions for the modified Ward identities for the vertex functions in a uniform field configuration $\Gamma^{(n,0)}(p_1,\dots,p_{n-1})$ and  $\Gamma^{(n,1)}(p_1,\dots,p_{n})$, the Fourier transforms of $\frac{\delta^{n}\Gamma}{\delta \varphi(x_1)\dots\delta \varphi(x_n)}$ fixing $x_n=0$ and $\frac{\delta^{n+1}\Gamma}{\delta \varphi(x_1)\dots\delta \varphi(x_n)\delta K(y_1)}$ fixing $y_1=0$, respectively. Notice that the freedom in fixing one coordinate to zero is due to translation invariance which is already implemented and imposes momentum conservation for the vertex functions. Following the convention of \cite{Cabrera2026}, we introduce the functions $H^{(n,0)}(q,p_1,\dots,p_{n-1},q')$ and $H^{(n,1)}(q,p_1,\dots,p_n,q')$ as the Fourier transforms of
\begin{equation}
\begin{split}
    H^{(n,0)}&(x,y,x_1,\dots,x_{n-1})=\\\int_{z,w}&G^{-1}(x,z)\frac{\delta^n G[z,w]}{\delta\varphi(x_1)\dots\delta\varphi(x_{n-1})\delta\varphi(0)}G^{-1}(w,y),
\end{split}
\end{equation} and 
\begin{equation}
\begin{split}
    H^{(n,1)}&(x,y,x_1,\dots,x_n)=\\\int_{z,w}&G^{-1}(x,z)\frac{\delta^{n+1} G[z,w]}{\delta\varphi(x_1)\dots\delta\varphi(x_n)\delta K(0)}G^{-1}(w,y),
\end{split}
\end{equation}
respectively, which include all the diagrammatic contributions involving vertex functions from $\Gamma^{(2,0)}$ up to $\Gamma^{(n+2,1)}$ \cite{Dupuis2021,Delamotte2021,Cabrera2025}. With this notation, and disregarding rotations, the modified Ward identities then take the form:

\begin{widetext}
\textit{Dilatations for $\Gamma^{(n,0)}$}:
\begin{equation}\label{dilGamman0}
    \Big[\big(\sum_{i=1}^{n-1}p_i^{\nu}\frac{\partial}{\partial p_i^{\nu}}\big)-d+nD_{\varphi}\vphantom{\sum_{i=1}^{n-1}}+D_{\varphi}\phi\frac{\partial}{\partial \phi}\Big]\Gamma^{(n,0)}(p_1,\dots,p_{n-1})=\frac{1}{2}\int_{q}\dot{R}_k(q)G^2(q)H^{(n,0)}(q,p_1,\dots,p_{n-1},-q).
\end{equation}

\textit{Dilatations for $\Gamma^{(n,1)}$}:
\begin{equation}\label{dilGamman1}
    \Big[\big(\sum_{i=1}^{n}p_i^{\nu}\frac{\partial}{\partial p_i^{\nu}}\big) -D_{\mathcal{O}}+nD_{\varphi}\vphantom{\sum_{i=1}^{n-1}}+D_{\varphi}\phi\frac{\partial}{\partial \phi}\Big]\Gamma^{(n,1)}(p_1,\dots,p_{n})=\frac{1}{2}\int_{q}\dot{R}_k(q)G^2(q)H^{(n,1)}(q,p_1,\dots,p_{n},-q).
\end{equation}

\textit{Special Conformal for $\Gamma^{(n,0)}$}:
\begin{equation}\label{confGamman0}
    \begin{split}
        \sum_{i=1}^{n-1}\Big[\vphantom{\sum_{i=1}^{n}}p_i^{\mu}\frac{\partial^2}{\partial p_i^{\nu}\partial p_i^{\nu}}-&2p_i^{\nu}\frac{\partial^2}{\partial p_i^{\nu}\partial p_i^{\mu}}
        -2D_{\varphi}\frac{\partial}{\partial p_i^{\mu}}\vphantom{\sum_{i=1}^{n}}\Big]\Gamma^{(n,0)}(p_1,\dots,p_{n-1})
        -2D_{\varphi}\phi \frac{\partial}{\partial r^{\mu}}\Gamma^{(n+1,0)}(r,p_1,\dots)\Big\vert_{r=0}= 
        \\
        -&\frac{1}{2}\int_{q}\dot{R}_k(q)G(q)\Big(\frac{\partial}{\partial q^{\mu}}+\frac{\partial}{\partial q'^{\mu}}\Big)\Big\lbrace H^{(n,0)}(q,p_1,\dots,p_{n-1},q')\Big\rbrace\Big\vert_{q'=-q}G(q).
    \end{split}
\end{equation}

\textit{Special Conformal for $\Gamma^{(n,1)}$}:
\begin{equation}\label{confGamman1}
    \begin{split}
        \sum_{i=1}^{n}\Big[\vphantom{\sum_{i=1}^{n}}p_i^{\mu}\frac{\partial^2}{\partial p_i^{\nu}\partial p_i^{\nu}}-&2p_i^{\nu}\frac{\partial^2}{\partial p_i^{\nu}\partial p_i^{\mu}}-2D_\varphi\frac{\partial}{\partial p_i^{\mu}}\vphantom{\sum_{i=1}^{n}}\Big]\Gamma^{(n,1)}(p_1,\dots,p_{n}) -2D_{\varphi}\phi \frac{\partial}{\partial r^{\mu}}\Gamma^{(n+1,1)}(r,p_1,\dots)\Big\vert_{r=0}=\\
        -&\frac{1}{2}\int_{q}\dot{R}_k(q)G(q)
        \Big(\frac{\partial}{\partial q^{\mu}}+\frac{\partial}{\partial q'^{\mu}}\Big)
        \Big\lbrace H^{(n,1)}(q,p_1,\dots,p_{n},q')\Big\rbrace\Big\vert_{q'=-q}G(q).
    \end{split}
\end{equation}
\end{widetext}

A crucial observation is that the Ward identities for dilatation \eqref{dilGamman0} or \eqref{dilGamman1} coincide with the fixed point equations \eqref{eq:FixPoint} when written for the vertex functions. In this sense, we can say that, at the fixed point, scale invariance is realized, even if approximations are performed. The situation is different for \eqref{confGamman0} and \eqref{confGamman1}  because the truncation scheme is, in general, not compatible with conformal invariance.

When including the source for composite operators, already at NLO of the DE (that is, order $O(\partial^2)$), the conformal Ward identity yields a single nontrivial constraint. This is at odds with the situation in the absence of the source $K$, where the first nontrivial constraint from conformal symmetry occurs when including NNLO terms.

In this work, we study the behavior of these constraints at the fixed point, as the regulating function varies. To this end, we vary the multiplicative factor $\alpha$, which has been shown to be very influential and decisive for computing universal critical quantities. We write these constraints as $\mathcal{C}^{(n,m)}_{i}(\rho,\alpha)$, where $\rho\equiv\varphi^2/2$, meaning that it is the $i$-th special conformal constraint obtained for the vertex function $\Gamma^{(n,m)}$.\footnote{The number of conformal constraints arising from a given vertex function is ultimately bounded by the maximum order of approximation considered in this work. This means that if we were to implement order $O(\partial^6)$, more constraints would arise from these vertex functions altogether with new ones for higher vertices.} When only one constraint is present, we omit the subscript $i$. These constraints are made of two parts, one corresponding to a dimensional expression which is the part corresponding to the Ward identity without regulator and that we shall denote as $\mathcal{C}^{(n,m)}_{i,L}$ while the part which belongs to the loop integral contribution which will be denoted as $\mathcal{C}^{(n,m)}_{i,R}$, as a reference to left-hand-side and right-hand-side of Eqs.~(\ref{confGamman0}-\ref{confGamman1}). The explicit expressions for these are given in the supplemental material.

As was the case in previous studies, we will study the conformal constraints by means of the following normalized function:
\begin{equation}\label{confConstDENorm}
	f^{(n,m)}_{i}(\rho,\alpha)=\Big[\mathcal{C}^{(n,m)}_{i,L}(\rho)-\mathcal{C}^{(n,m)}_{i,R}(\rho)\Big]\Big(1+\frac{\rho}{\rho_0}\Big)^{-\chi},
\end{equation} 
where $\chi$ is the scaling dimension of the corresponding dimensional term and $\rho_0$ is the value of $\rho$ corresponding to the minimum of the potential, which plays the role of a reference value to avoid affecting the small $\rho$ behavior. The reason to consider this combination, as explained in \cite{Delamotte2021,Cabrera2025}, is due to the fact that at large $\rho$, the asymptotic behavior of the constraint is governed only by the dimensional part denoted with the subscript $L$ and, consequently, it may arbitrarily diverge or vanish depending on the relevance of the composite operator under consideration.

In practice, these $f^{(n,m)}_{i}$ constraints reduce to a specific combination of the ansatz functions and its derivatives. They provide a quantitative measure of conformal symmetry breaking within the truncation.
We implement the PMC criterion based on the behavior of the functions $f^{(n,m)}_{i}$. In previous works \cite{Balog2020,Delamotte2021,Cabrera2025}, it was used that the parameter $\alpha$ should be fixed at $\alpha_{\mathrm{PMC},0}$ so that $|f^{(n,m)}_{i}(\rho=0,\alpha)|$ is minimal:
\begin{equation}\label{eq:PMC0}
    \alpha_{\mathrm{PMC},0}
    =
    \operatorname*{arg\,min}_{\alpha}
    \left|f_i^{(n,m)}(0,\alpha)\right|.
\end{equation}
This seemingly arbitrary criterion was justified by the observation that different implementations tend to produce similar results, with discrepancies well within the error bars. However, as we will show in Sec.~\ref{Sec:Res}, a different criterion here may prove useful. In this work, we summarize at the end the results corresponding to the criterion yielding $\alpha_{PMC,0}$, but we present mainly results corresponding to the fixing of the parameter $\alpha$ at the value $\alpha_{PMC,||\cdot||}$ given by: 
\begin{equation}\label{eq:PMCnorm}
    \alpha_{\mathrm{PMC},\|\cdot\|}
    =
    \operatorname*{arg\,min}_{\alpha}
    \left[
        \int_{0}^{\rho_{\max}}
        \left|f_i^{(n,m)}(\rho,\alpha)\right|^2
        \,d\rho
    \right]^{1/2},
\end{equation}
where $\rho_{\max}$ denotes the upper bound of the field range considered. This criterion therefore selects the value of $\alpha$ that minimizes the violation over the whole field range under study. 
At this point, we would like to call the attention to the fact that, even though this criterion tends to prioritize larger field values, this, in fact, is not a drawback. This is due to the fact that at large fields, the constraint is dominated by its scaling part (what we have called $\mathcal{C}_{i,L}^{(n,m)}$) which is an exact relation that must hold for the functions of the ansatz. In this sense, this criterion interpolates between considering only the asymptotic case of large fields (which was already mentioned in \cite{Cabrera2025}) and the one given by Eq.~\eqref{eq:PMC0}.

\section{Results}\label{Sec:Res}

We now present the study of conformal constraints appearing at order $O(\partial^4)$ of the DE in the presence of a source for composite operators, within the truncation given by \eqref{Eq:ansatz}.

Among the conformal constraints that arise at this truncation, six are independent and genuinely distinct from those derived from scale invariance. They are obtained as follows: a) pick a constraint $\mathcal C^{(n,m)}$ obtained by applying the conformal Ward Identity to the vertex $\Gamma^{(n,m)}$ and b) extract the part proportional to some momentum configuration. In Table \ref{tab:constraints}, we describe how these constraints are obtained exactly. 
 \begin{table}[htbp]
     \centering
     \begin{tabular}{c|c|c}
          Constraint& Vertex&Momentum structure  \\
          \hline
          $\mathcal C^{(3,0)}$ & $\Gamma^{(3,0)}$ & $p_{1}^\mu p_2^2$ \\
          $\mathcal C^{(1,1)}_1$ & $\Gamma^{(1,1)}$ & $p^\mu$  \\
          $\mathcal C^{(1,1)}_2$ & $\Gamma^{(1,1)}$ & $p^\mu p^2$ \\
          $\mathcal C^{(2,1)}_1$ & $\Gamma^{(2,1)}$ & $p_{1}^\mu p_2^2$ \\
          $\mathcal C^{(2,1)}_2$ & $\Gamma^{(2,1)}$ & $p_{1}^\mu p_1\cdot p_2$ \\
          $\mathcal C^{(3,1)}$ & $\Gamma^{(3,1)}$ & $p_{1}^\mu p_2\cdot p_3$ \\
     \end{tabular}
     \caption{Characterization of the various conformal constraints obtained at order $\partial^4$, for the ansatz \eqref{Eq:ansatz}. }
     \label{tab:constraints}
 \end{table}
 One of these constraints, $\mathcal C_1^{(1,1)}$, appears at order $O(\partial^2)$ and receives corrections at order $O(\partial^4)$, being accessible at two different orders of the DE. Whenever needed, we will explicitly indicate the order under consideration. The other five constraints require considering at least $O(\partial^4)$ truncations.

In previous works, the only constraint stemming from $\Gamma_k^{(3,0)}$ was studied at this order \cite{Balog2020} and the constraint $\mathcal{C}^{(1,1)}_{1}$ stemming from $\Gamma_k^{(1,1)}$ at order $O(\partial^2)$ was studied in another work \cite{Delamotte2021} (it is worth mentioning that this constraint was also studied for $O(N)$ models in \cite{Cabrera2025}). This means that this is the first opportunity to study a particular constraint at successive orders of the DE, which will allow us to assess the convergence of the approximation scheme from the conformal symmetry point of view.

We remark that, as evidenced in \cite{Delamotte2021,Cabrera2025}, for the perturbation associated with the critical exponent $\omega_2$, conformal symmetry seems to behave differently than for the less irrelevant operator associated with $\omega$ and for the relevant operator associated with $\nu$. In this study, we found the same and even worsened behavior when considering different conformal constraints. Moreover, given the fact that $\omega_2$ becomes complex valued and departs from its expected value, this probably implies that the DE approach is not being able to correctly resolve the $\omega_2$ sector.  Finally, we also mention that for the operator associated with the next irrelevant exponent, say $\omega_3$, the behavior of conformal constraints is once again similar to the cases of $\nu$ and $\omega$ and, moreover, the exponent is fully real and well differentiated from other exponents. For the reasons just described, we won't discuss in this work the perturbation associated with critical exponent $\omega_2$ or more irrelevant ones.

\subsection{Convergence of the derivative expansion at successive orders: the $\mathcal{C}^{(1,1)}_{1}$ constraint}

\begin{figure}[tpb]
    \centering
    \begin{subfigure}{0.45\textwidth}
        \centering
        \includegraphics[width=\linewidth]{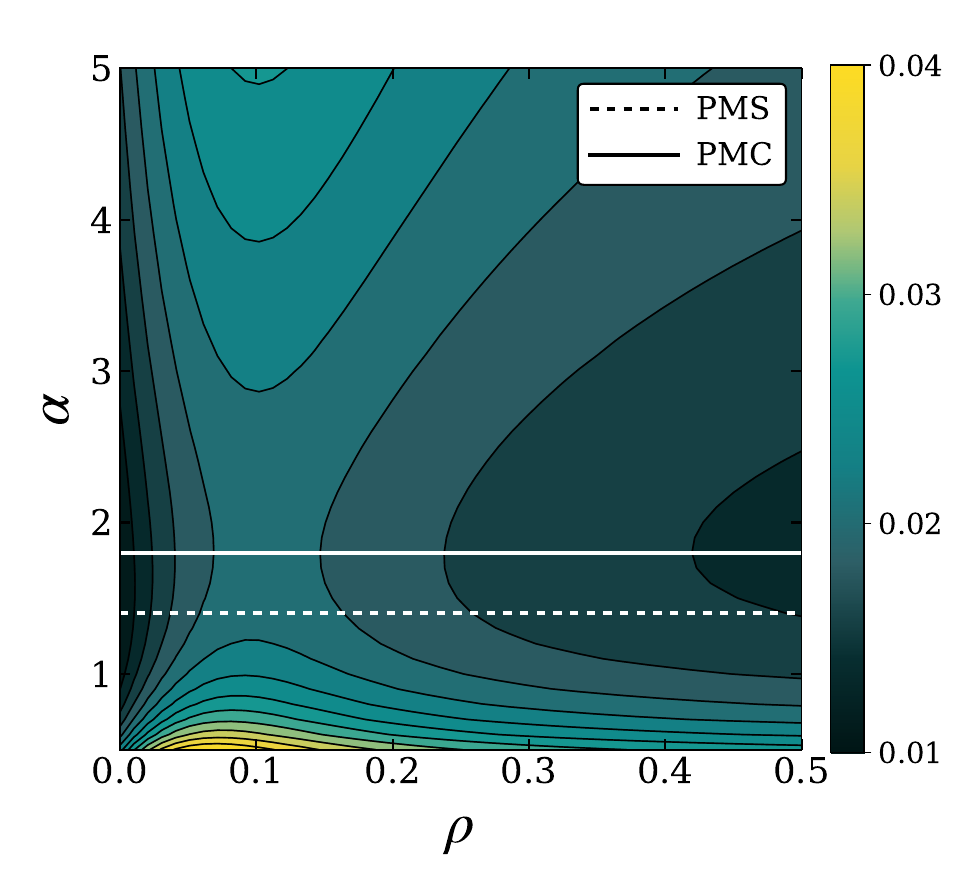}
        \caption{Perturbation associated with critical exponent $\nu$.}
        \label{fig:f11de2nu}
    \end{subfigure}
    \hfill
    \begin{subfigure}{0.45\textwidth}
        \centering
        \includegraphics[width=\linewidth]{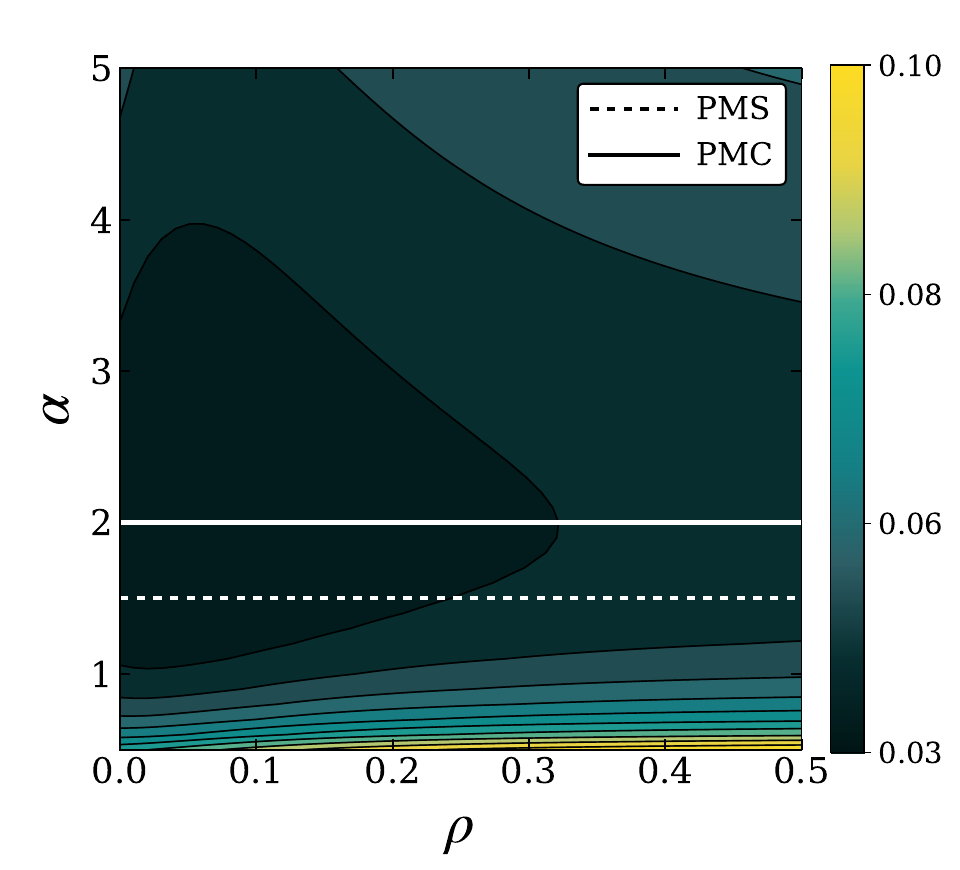}
        \caption{Perturbation associated with critical exponent $\omega$.}
        \label{fig:f11de2omega}
    \end{subfigure}
    \caption{Function $|f^{(1,1)}_{1,\partial^2}(\rho,\alpha)|$ corresponding to perturbations associated to critical exponents $\nu$ and $\omega$ using the exponential regulator given in \eqref{eq:regulator-exp} at order $O(\partial^2)$ of the DE. Horizontal lines correspond to $\alpha_{\text{PMS}}$ (dashed) and $\alpha_{\text{PMC}}$ (continuous).}
    \label{fig:f11de2}
\end{figure}

We start the presentation of results by discussing the constraint $\mathcal{C}^{(1,1)}_{1}$. As previously mentioned, the constraint was analyzed at order $O(\partial^2)$ for composite operators associated with critical exponents $\nu$ and $\omega$ \cite{Delamotte2021}. It was evidenced from this that the violation is less marked at regulator profiles close to the PMS criterion for $\nu$ and $\omega$ operators. In Fig.~\ref{fig:f11de2}, we reproduce the results for $f^{(1,1)}_{1,\partial^2}(\rho,\alpha)$ that were previously obtained for operators associated with $\nu$ and $\omega$ exhibiting the described behavior. In these figures, PMS and PMC values for the associated critical exponents are marked with straight lines (dashed and solid, respectively).

\begin{figure}[tpb]
    \centering
    \begin{subfigure}{0.45\textwidth}
        \centering
        \includegraphics[width=\linewidth]{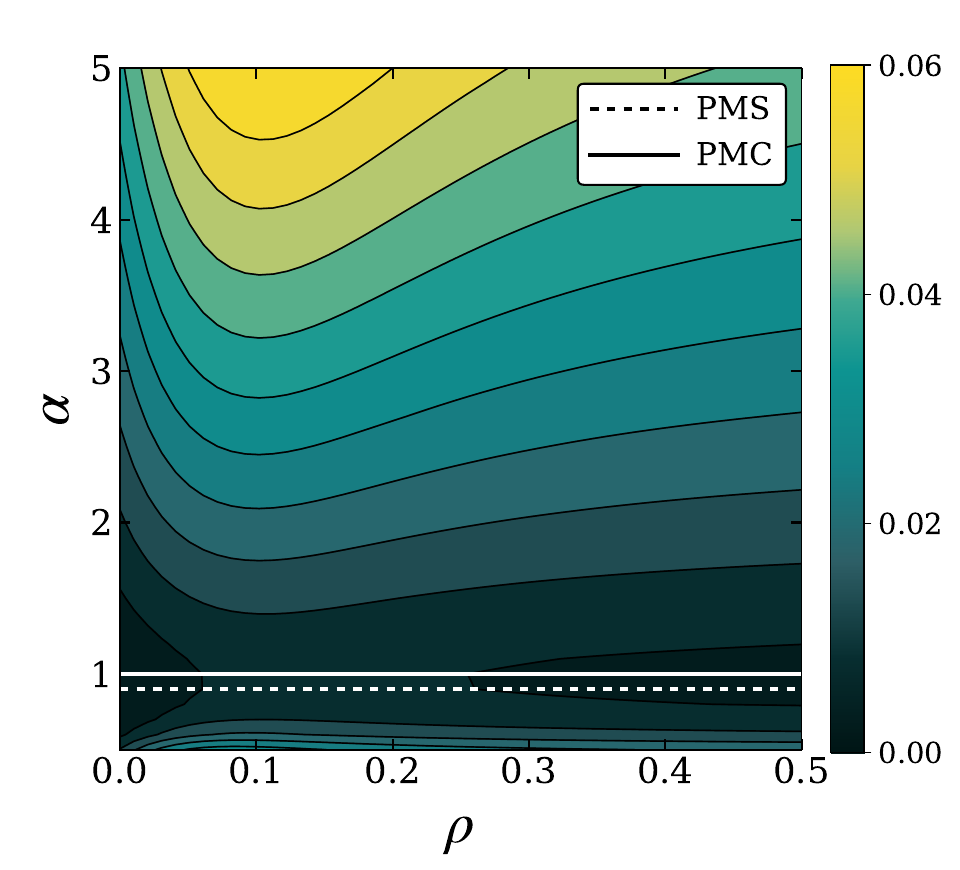}
        \caption{Perturbation associated with critical exponent $\nu$.}
        \label{fig:f11de4nu}
    \end{subfigure}
    \hfill
    \begin{subfigure}{0.45\textwidth}
        \centering
        \includegraphics[width=\linewidth]{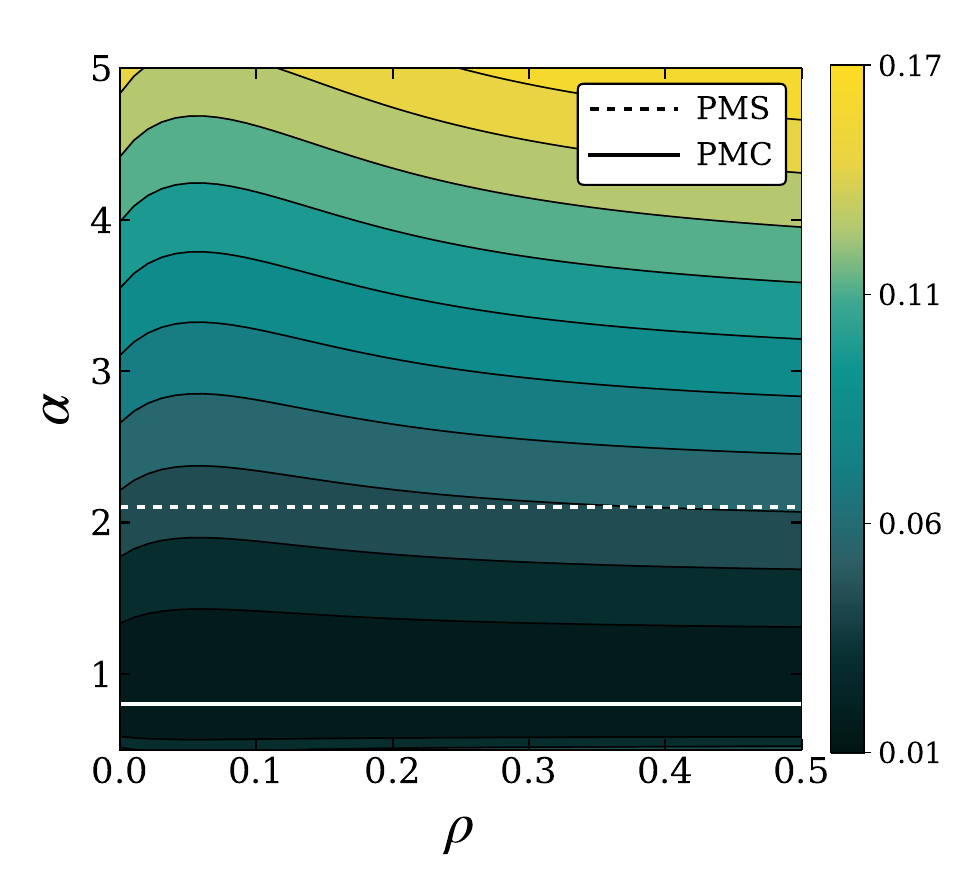}
        \caption{Perturbation associated with critical exponent $\omega$.}
        \label{fig:f11de4omega}
    \end{subfigure}
    \caption{Function $|f^{(1,1)}_{1,\partial^4}(\rho,\alpha)|$ corresponding to perturbations associated to critical exponents $\nu$ and $\omega$ using the exponential regulator given in \eqref{eq:regulator-exp} at order $O(\partial^4)$ of the DE. Horizontal lines correspond to $\alpha_{\text{PMS}}$ (dashed) and $\alpha_{\text{PMC}}$ (continuous) whenever available.}
    \label{fig:f11de4}
\end{figure}

When considering the next order of approximation, namely $O(\partial^4)$, we find a similar qualitative behavior for the perturbations associated with the critical exponents $\nu$ and $\omega$. Figure~\ref{fig:f11de4} displays the corresponding behavior of $f^{(1,1)}_{1,\partial^4}(\rho,\alpha)$.
To better analyze the constraints at successive orders, it is useful to consider the ratio between two consecutive orders, defined as 
\begin{equation}\label{eq:ratiof11}
g(\rho,\bar{\alpha})\equiv\frac{|f^{(1,1)}_{1,\partial^4}(\rho,\bar{\alpha}\alpha^{(4)}_{PMC,\|\cdot\|})|}{|f^{(1,1)}_{1,\partial^2}(\rho,\bar{\alpha}\alpha^{(2)}_{PMC,\|\cdot\|})|}.
\end{equation}
where $\bar{\alpha}$ denotes the common normalized regulator parameter, $\bar{\alpha}=\alpha^{(2)}/\alpha^{(2)}_{PMC,\|\cdot\|}=\alpha^{(4)}/\alpha^{(4)}_{PMC,\|\cdot\|}$.

\begin{figure}[tpb]
    \centering
    \begin{subfigure}{0.45\textwidth}
        \centering
        \includegraphics[width=\linewidth]{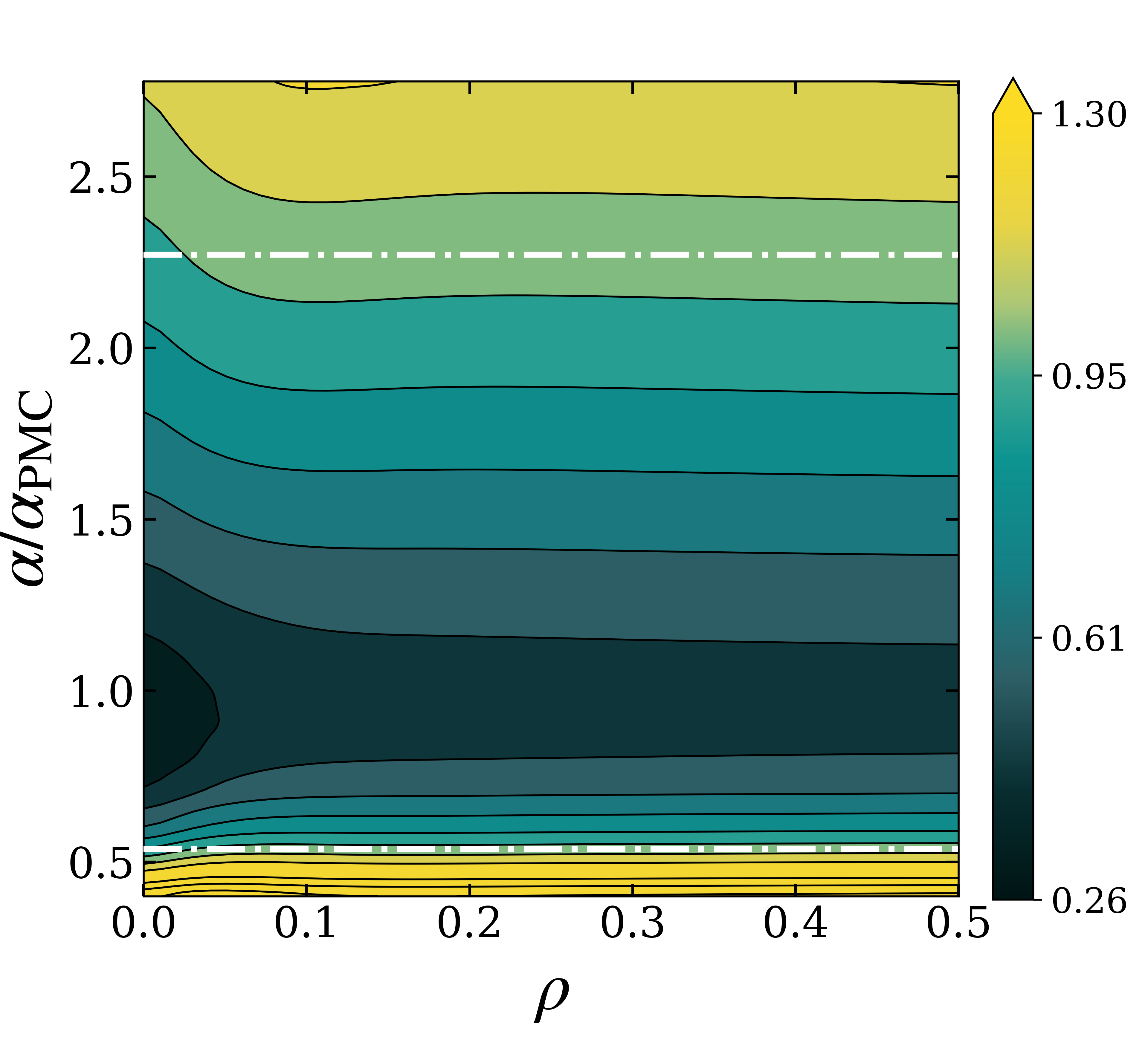}
        \caption{Perturbation associated with critical exponent $\nu$.}
        \label{fig:g11nu}
    \end{subfigure}
    \hfill
    \begin{subfigure}{0.45\textwidth}
        \centering
        \includegraphics[width=\linewidth]{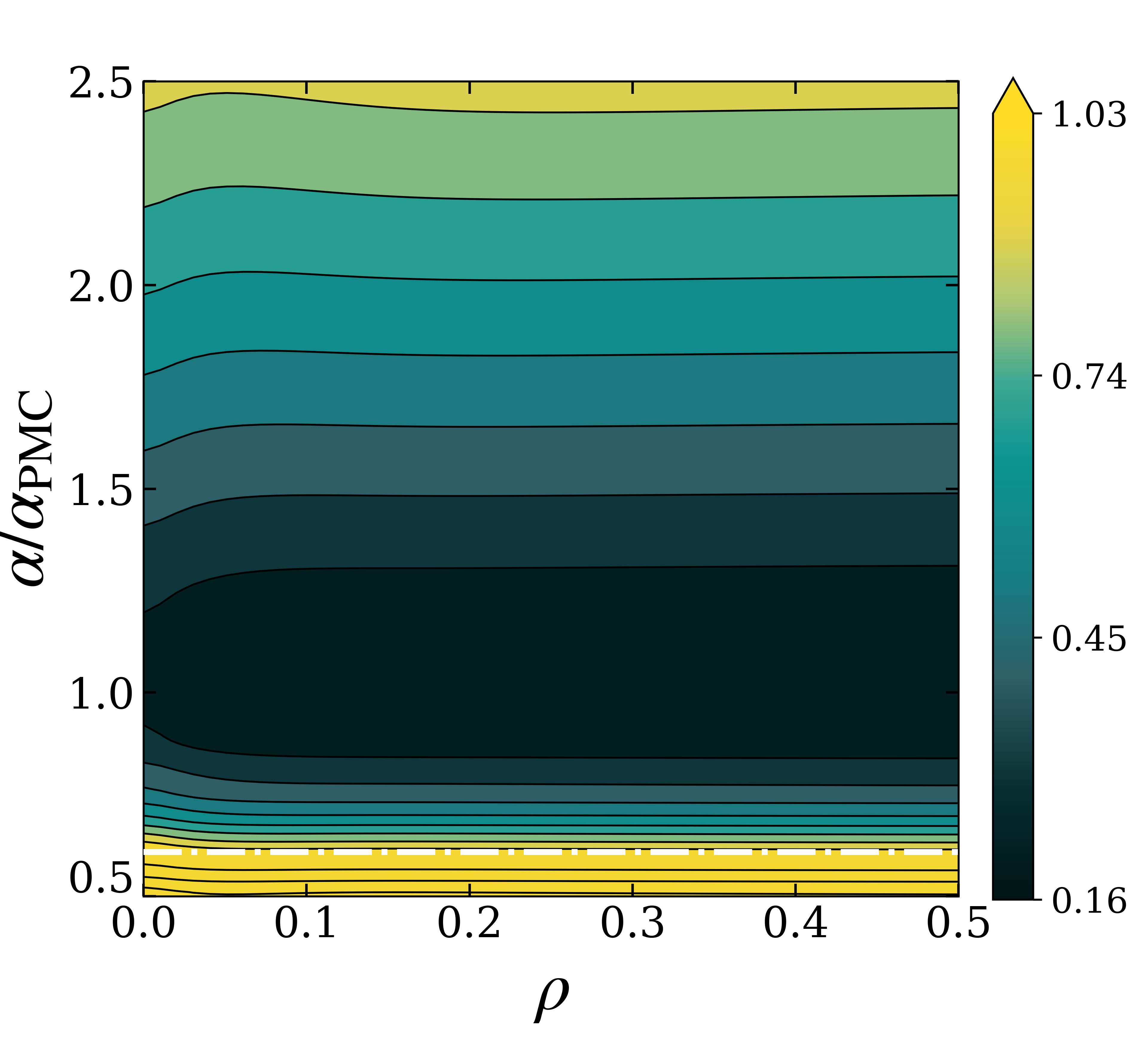}
        \caption{Perturbation associated with critical exponent $\omega$.}
        \label{fig:g11omega}
    \end{subfigure}
    \caption{Function $g(\rho,\bar{\alpha})$ corresponding to perturbations associated with the critical exponents $\nu$ and $\omega$, using the exponential regulator given in \eqref{eq:regulator-exp}. White dot-dashed lines indicate the values of $\bar{\alpha}=\alpha/\alpha_{\mathrm{PMC}}$ for which $\|g(\cdot,\bar{\alpha})\|=1$.}
    \label{fig:g11}
\end{figure}

We show the behavior of the function $g(\rho,\bar{\alpha})$ given in Eq.~\eqref{eq:ratiof11} in Fig.~\ref{fig:g11}, associated with critical exponents $\nu$ and $\omega$, respectively. We highlight the fact that there exists a range of $\bar{\alpha}$ for which the constraint at order $O(\partial^4)$ improves with respect to the one at order $O(\partial^2)$. These are indicated by dot-dashed lines for reference. Moreover, the smallest value of this ratio is around $1/4$, as expected from arguments put forward in \cite{Balog2019,DePolsi2020a,DePolsi2022}, which correlate this improvement with a more accurate determination of the momentum dependence of the two-point function. 
This demonstrates that incorporating $O(\partial^4)$ operators systematically improves the realization of conformal symmetry and thereby supports the convergence of the DE. 

\begin{figure}[tpb]
    \centering
    \begin{subfigure}{0.45\textwidth}
        \centering
        \includegraphics[width=\linewidth]{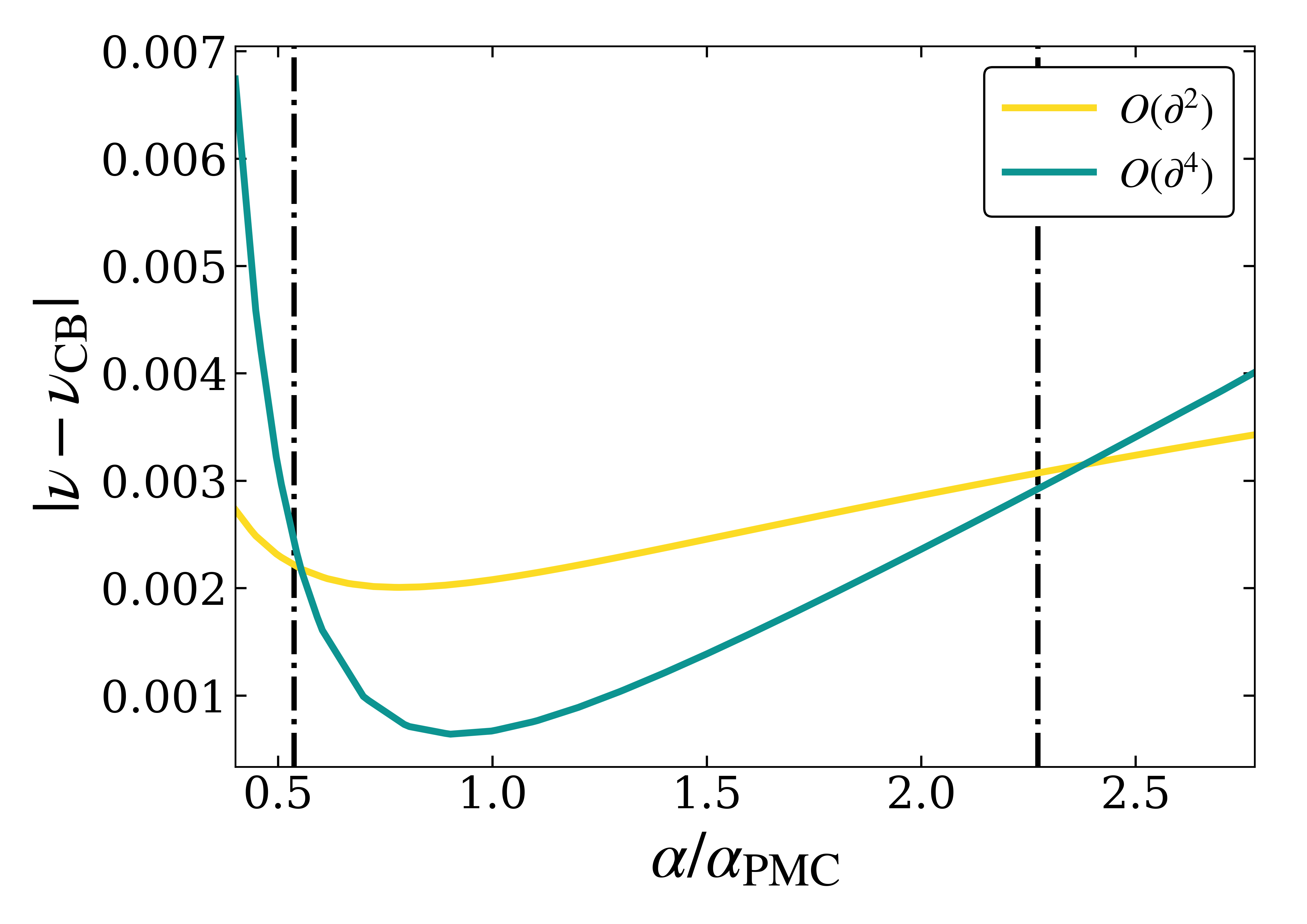}
        \caption{Critical exponent $\nu$.}
        \label{fig:nuvsalpha}
    \end{subfigure}
    \hfill
    \begin{subfigure}{0.45\textwidth}
        \centering
        \includegraphics[width=\linewidth]{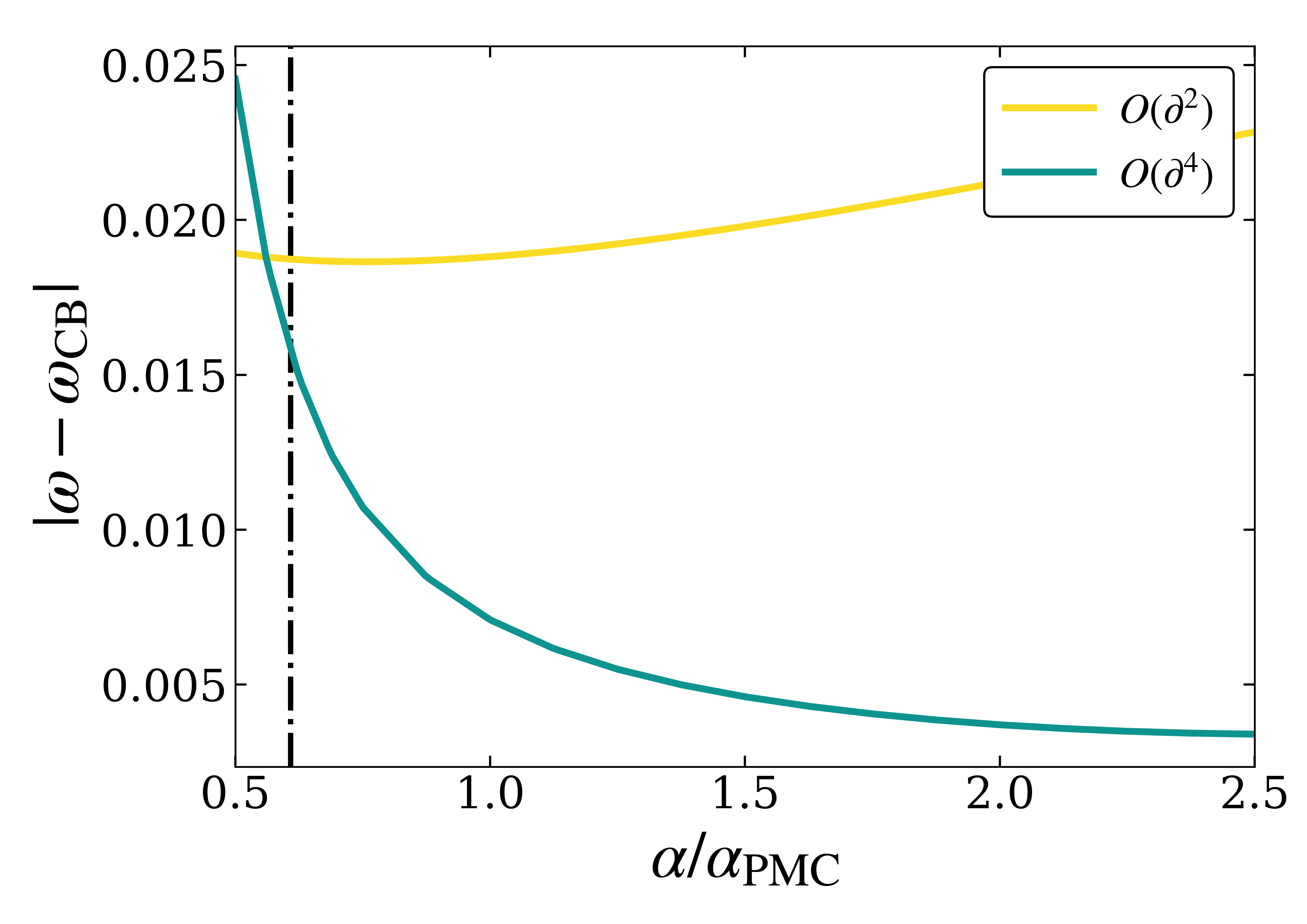}
        \caption{Critical exponent $\omega$.}
        \label{fig:omegavsalpha}
    \end{subfigure}
    \caption{Deviations of the critical exponents $\nu$ and $\omega$ from the reference values $\nu_{\mathrm{CB}}$ and $\omega_{\mathrm{CB}}$, respectively, obtained from the conformal bootstrap \cite{ElShowk2012}, as functions of $\alpha/\alpha_{\mathrm{PMC}}$, at orders $O(\partial^2)$ (yellow) and $O(\partial^4)$ (blue), using the exponential regulator given in \eqref{eq:regulator-exp}. Black dot-dashed lines mark the values of $\alpha/\alpha_{\mathrm{PMC}}$ for which $\|g(\cdot,\alpha)\|=1$.}
    \label{fig:critExpnuomde2and4}
\end{figure}

Outside this range of $\alpha$, not only is conformal symmetry broken even more, but this also coincides with a worsening of critical exponent predictions, as shown in Fig.~\ref{fig:critExpnuomde2and4}.

\subsection{New constraints of conformal symmetry at order $O(\partial^4)$ of the DE}

\begin{figure*}[t]
    \centering
    \begin{subfigure}{0.45\textwidth}
        \centering
        \includegraphics[width=\linewidth]{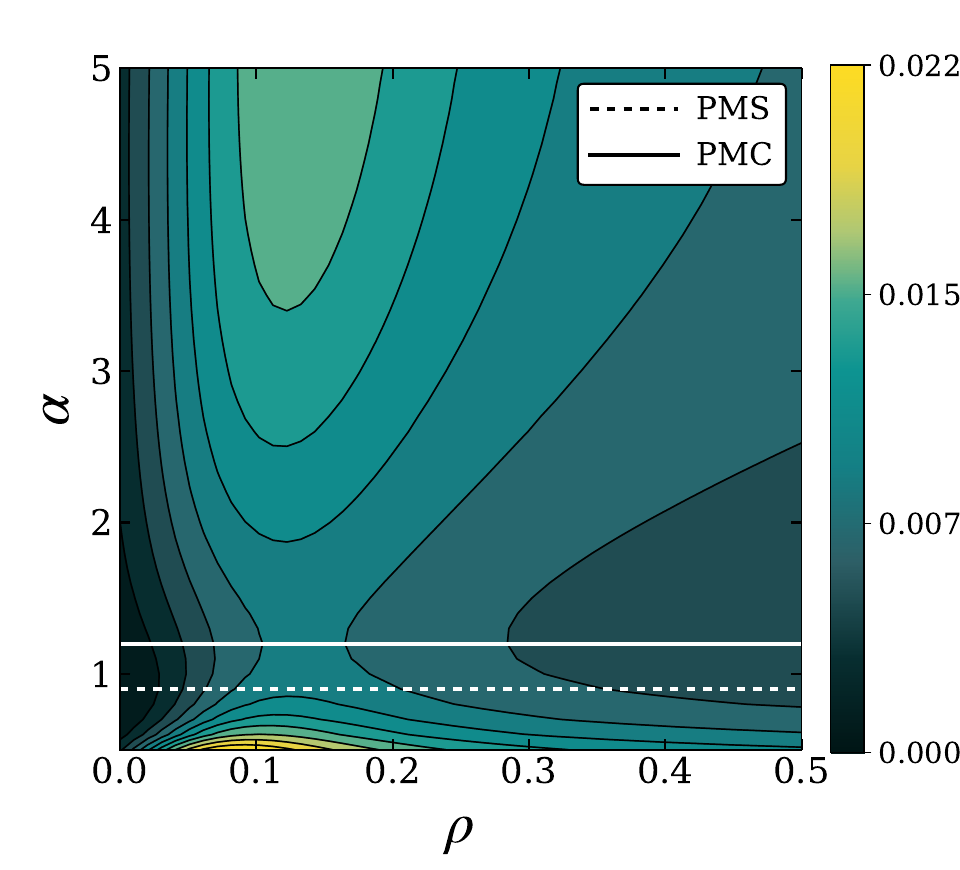}
        \caption{Violation of conformal symmetry constraint through function $|f^{(1,1)}_{2}(\rho,\alpha)|$.}
        \label{fig:5f112nu}
    \end{subfigure}
    \vspace{1em} 
    \hfill
    \begin{subfigure}{0.45\textwidth}
        \centering
        \includegraphics[width=\linewidth]{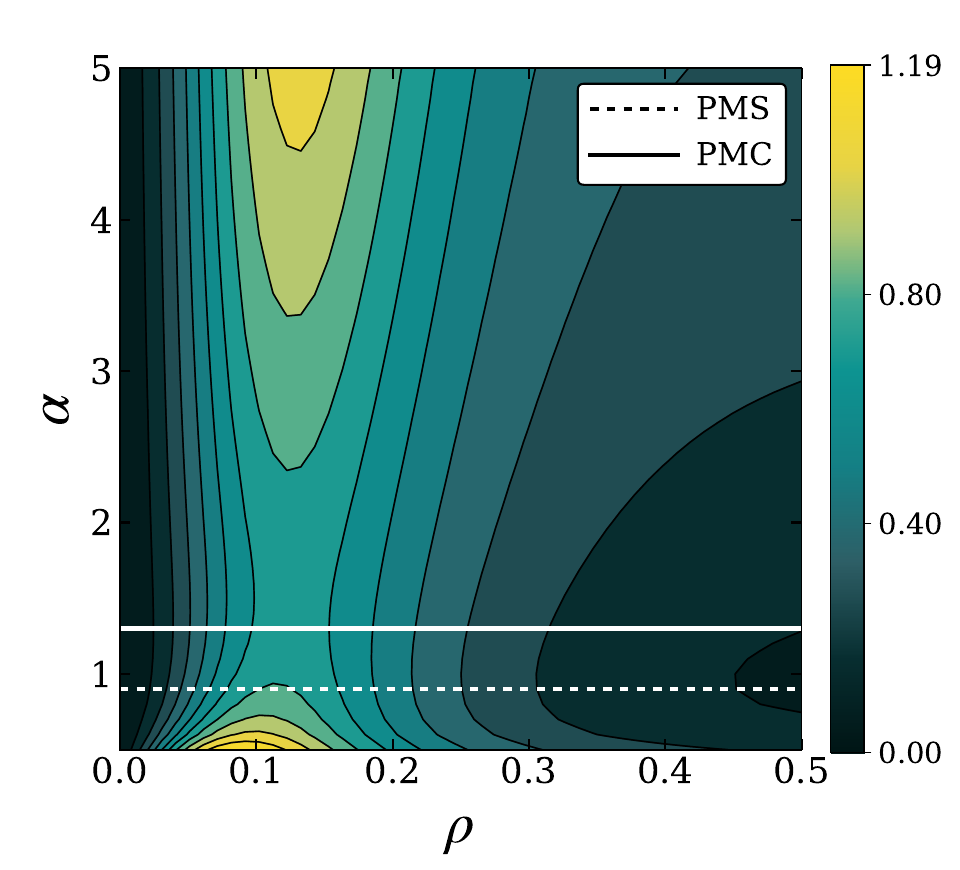}
        \caption{Violation of conformal symmetry constraint through function $|f^{(2,1)}_{1}(\rho,\alpha)|$.}
        \label{fig:5f211nu}
    \end{subfigure}
    \vspace{1em} 
    \begin{subfigure}{0.45\textwidth}
        \centering
        \includegraphics[width=\linewidth]{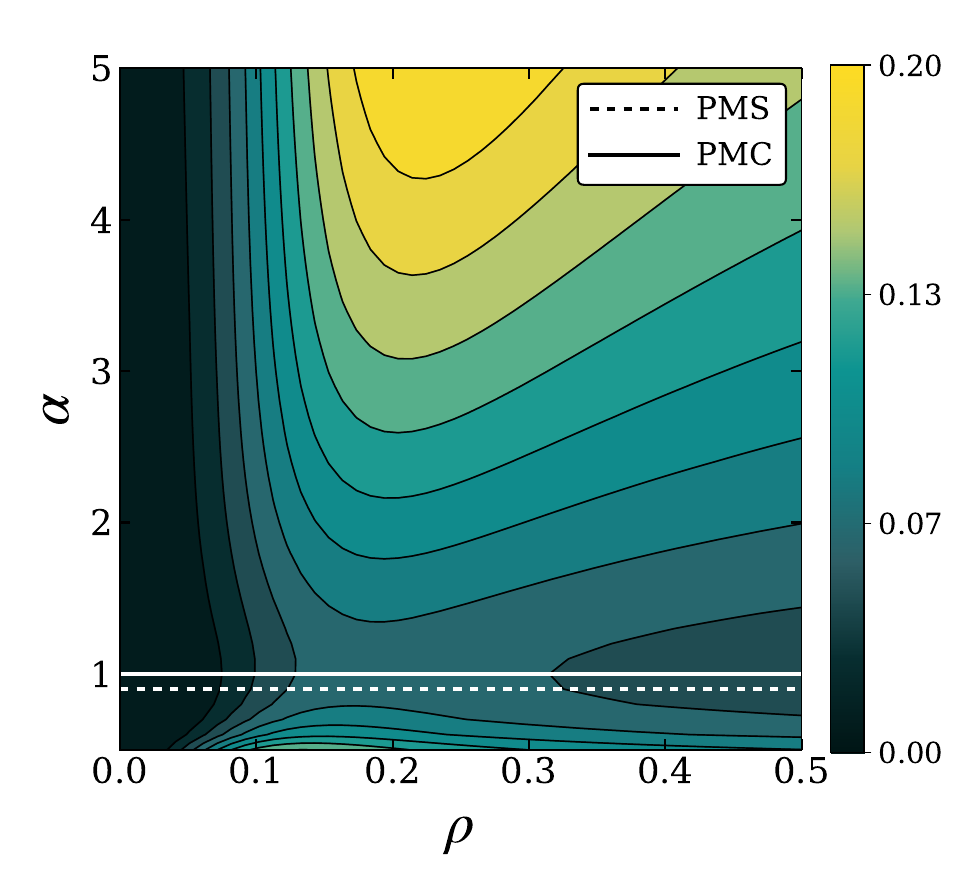}
        \caption{Violation of conformal symmetry constraint through function $|f^{(2,1)}_{2}(\rho,\alpha)|$.}
        \label{fig:5f212nu}
    \end{subfigure}
    \hfill
    \begin{subfigure}{0.45\textwidth}
        \centering
        \includegraphics[width=\linewidth]{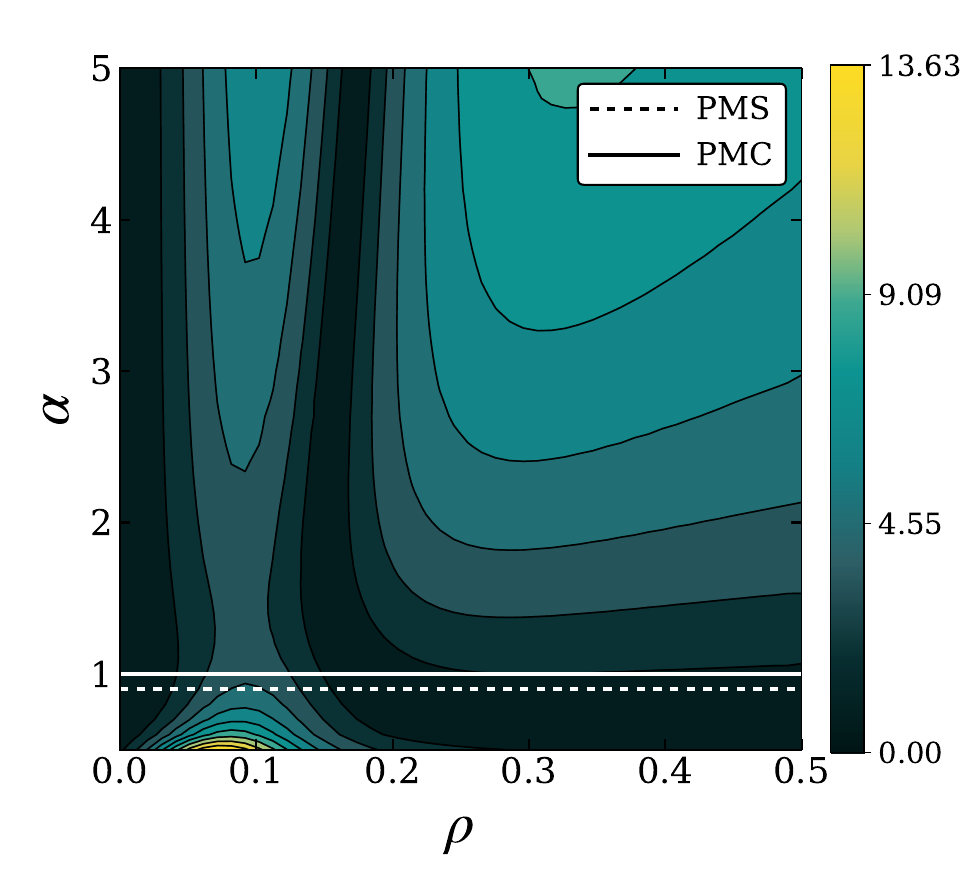}
        \caption{Violation of conformal symmetry constraint through function $|f^{(3,1)}(\rho,\alpha)|$.}
        \label{fig:5f31nu}
    \end{subfigure}
    \caption{Remaining conformal constraints study of function $|f^{(n,1)}_{i,\partial^4}(\rho,\alpha)|$ for the operator associated with critical exponent $\nu$ using the exponential regulator, Eq.~\eqref{eq:regulator-exp}.}
    \label{fig:5fnu}
\end{figure*}

\begin{figure*}[t]
    \centering
    \begin{subfigure}{0.45\textwidth}
        \centering
        \includegraphics[width=\linewidth]{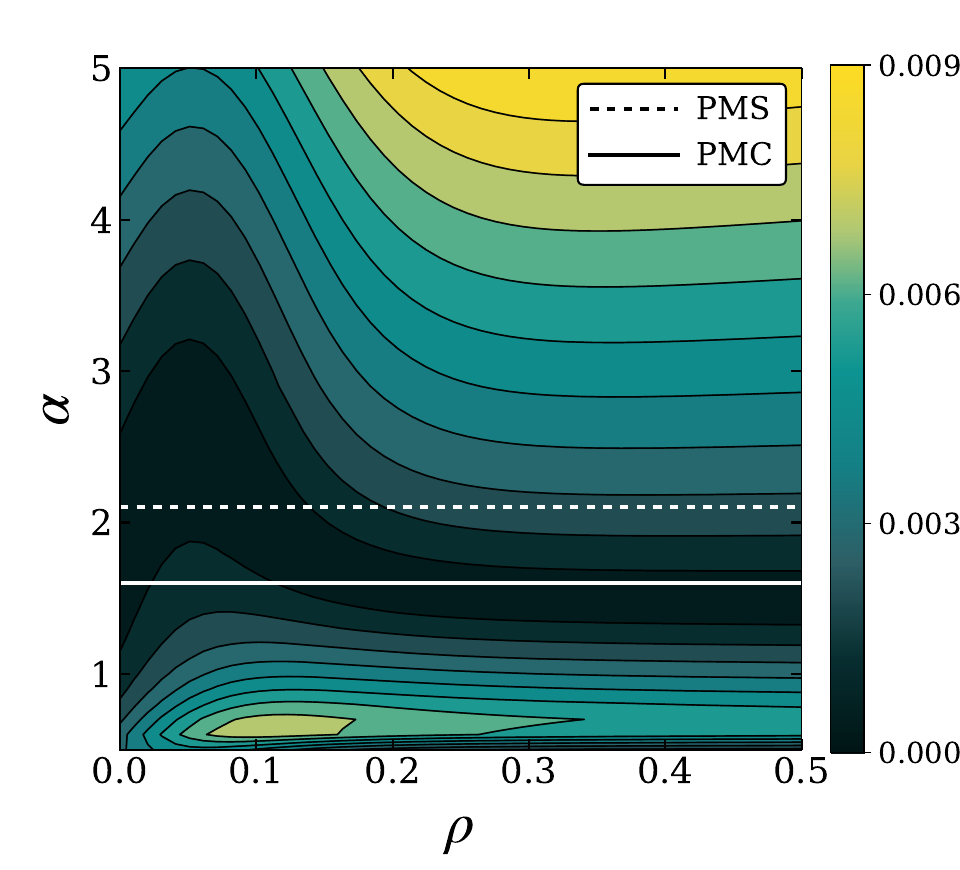}
        \caption{Violation of conformal symmetry constraint through function $|f^{(1,1)}_{2}(\rho,\alpha)|$.}
        \label{fig:5f112omega}
    \end{subfigure}
    \vspace{1em} 
    \hfill
    \begin{subfigure}{0.45\textwidth}
        \centering
        \includegraphics[width=\linewidth]{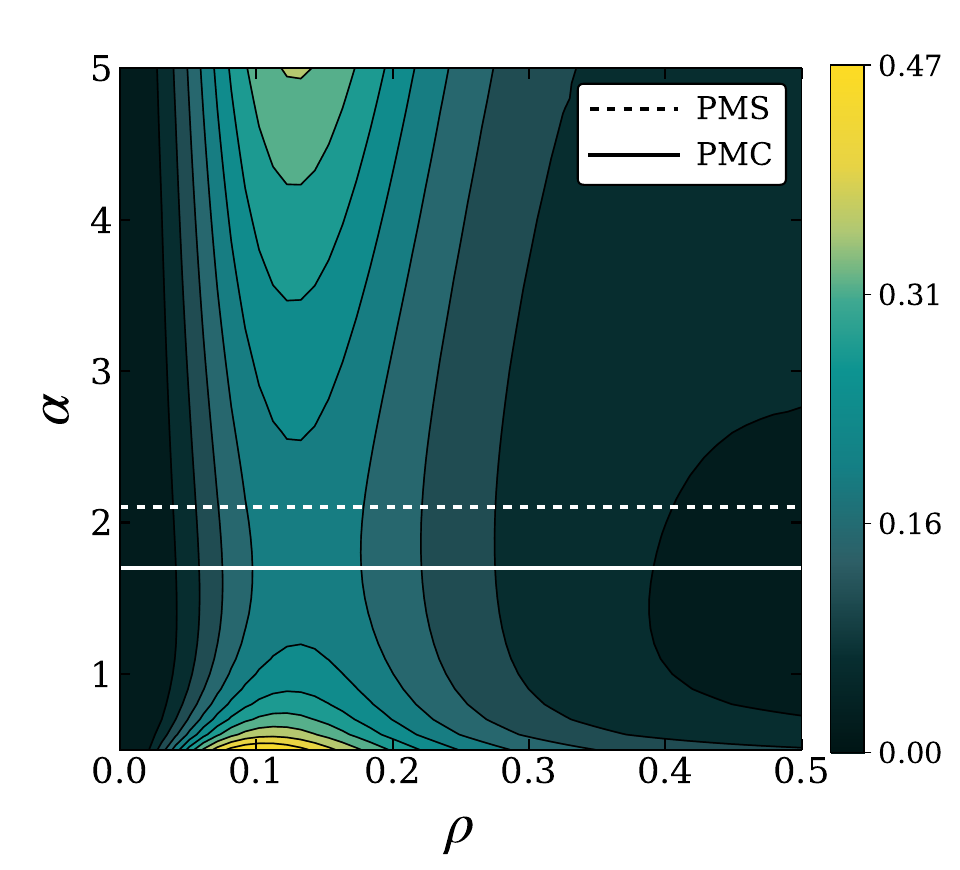}
        \caption{Violation of conformal symmetry constraint through function $|f^{(2,1)}_{1}(\rho,\alpha)|$.}
        \label{fig:5f211omega}
    \end{subfigure}
    \vspace{1em} 
    \begin{subfigure}{0.45\textwidth}
        \centering
        \includegraphics[width=\linewidth]{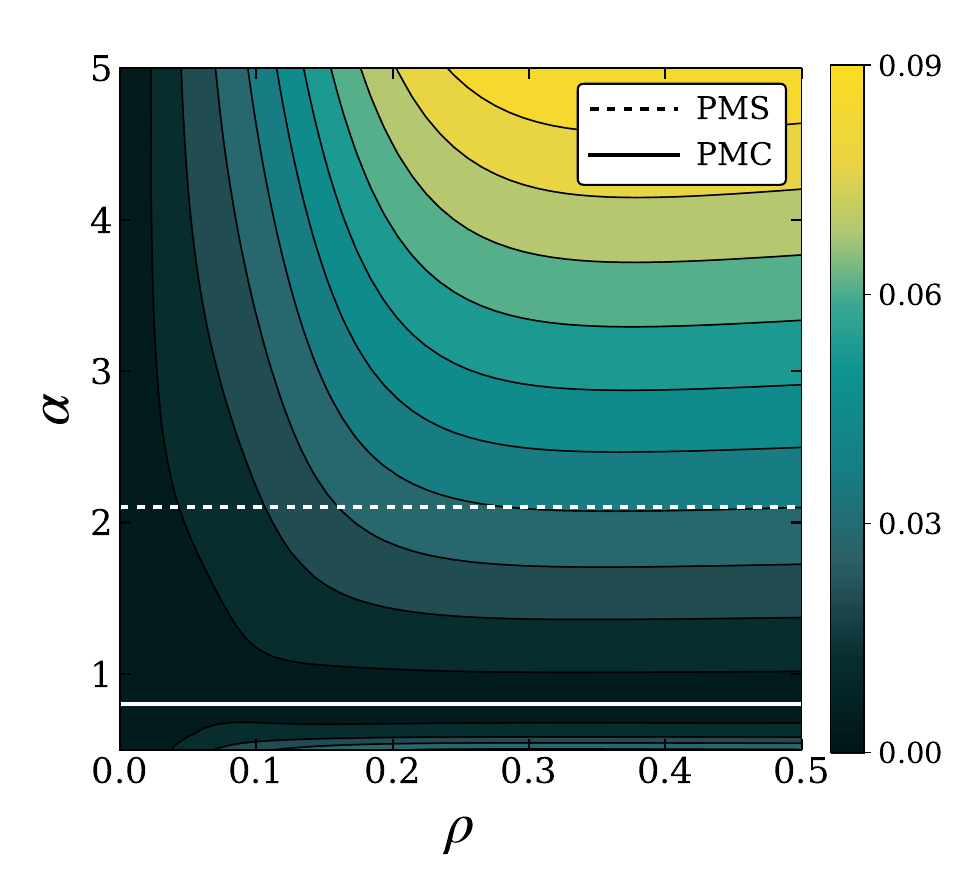}
        \caption{Violation of conformal symmetry constraint through function $|f^{(2,1)}_{2}(\rho,\alpha)|$.}
        \label{fig:5f212omega}
    \end{subfigure}
    \hfill
    \begin{subfigure}{0.45\textwidth}
        \centering
        \includegraphics[width=\linewidth]{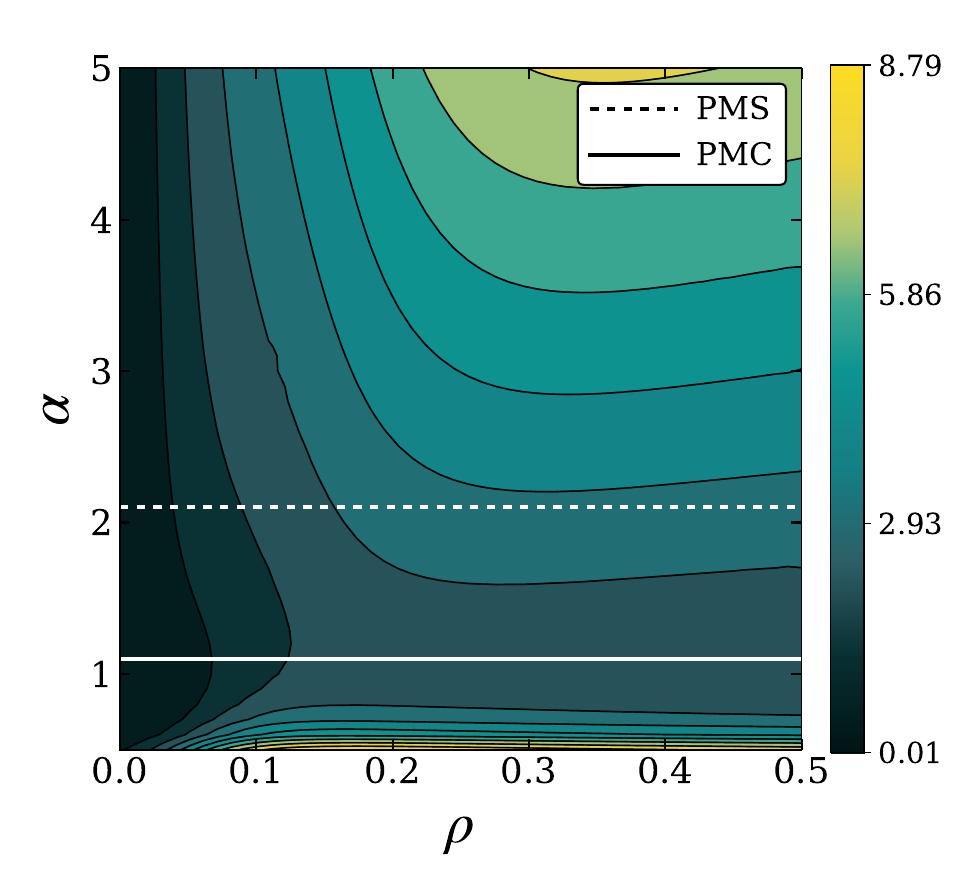}
        \caption{Violation of conformal symmetry constraint through function $|f^{(3,1)}(\rho,\alpha)|$.}
        \label{fig:5f31omega}
    \end{subfigure}
    \caption{Remaining conformal constraints study of function $|f^{(n,1)}_{i,\partial^4}(\rho,\alpha)|$ for the operator associated with critical exponent $\omega$ using the exponential regulator, Eq.~\eqref{eq:regulator-exp}.}
    \label{fig:5fomega}
\end{figure*}

Among the five remaining constraints, $\mathcal C^{(3,0)}$ was already studied in \cite{Balog2020} and there is nothing new we can say about it. We omit this one in our analysis and focus on the remaining four, which are shown in figures \ref{fig:5fnu} and \ref{fig:5fomega} for operators associated with critical exponents $\nu$ and $\omega$, respectively. Once again, we include for reference the PMS values of $\alpha$ for each of these critical exponents and the PMC values for $\alpha$ with straight lines.

From these figures, it is evidenced that the violations to constraints for operators associated with critical exponents $\nu$ and $\omega$ decrease in regions of $\alpha$ close to where critical exponents show minimal sensitivity. This implies that the extrema of physical observables and the minima of conformal constraint violations coincide within numerical accuracy. However, it must be emphasized that for some of the constraints the criterion of fixing $\alpha$ with the lesser violation of conformal symmetry at $\rho=0$ does not coincide with the clear overall behavior. This, in fact, happens due to a negligible violation for most $\alpha$ values at $\rho=0$ combined with the apparition of a global minimum at high $\alpha$ values (although the general behavior at non-zero $\rho$ is not compatible with this). In Tab.~\ref{tab:critExpExp} we summarize the obtained results for critical exponents using both PMC criteria discussed in Sec.~\ref{Sec:Symm} for each of these constraints and compare them to the predicted values for these exponents using the DE supplemented with the PMS criterion.

\begin{table*}[t]
\caption{Results obtained using the exponential regulator given in Eq.~\eqref{eq:regulator-exp} for critical exponents using different criteria for fixing the regulator parameter $\alpha$. A dash indicates that the corresponding PMC criterion yields no value for the highest-order vertex. The error bar for PMS data is calculated just using the difference between the raw results for this regulator at order $O(\partial^4)$ and at order $O(\partial^2)$ divided by 4 (see \cite{DePolsi2020a}).}\label{tab:critExpExp}
    \begin{ruledtabular}
        \begin{tabular}{lclccccc}
         \textbf{} & \textbf{$\partial^4$} & & \textbf{$\mathcal{C}^{(1,1)}_{1}$} & \textbf{$\mathcal{C}^{(1,1)}_{2}$} & \textbf{$\mathcal{C}^{(2,1)}_{1}$} & \textbf{$\mathcal{C}^{(2,1)}_{2}$} & \textbf{$\mathcal{C}^{(3,1)}$} \\
         \colrule
         \textbf{$\nu$} & 0.63061(66) & \textbf{$\nu$, $\alpha_{PMC,0}$} & 0.63061 & 0.63061 & 0.63086 & 0.63193 & - \\
         & &  \textbf{$\nu$, $\alpha_{PMC,\|\cdot\|}$} &  0.63064 & 0.63086 & 0.63101 & 0.63064 & 0.63064 \\
         \hline
         \textbf{$\omega$} & 0.8263(55) & \textbf{$\omega$, $\alpha_{PMC,0}$}  & 0.82259 & 0.82625 & 0.81665 & 0.82340 & 0.82582 \\
         & &  \textbf{$\omega$, $\alpha_{PMC,\|\cdot\|}$} & 0.82259 & 0.82598 & 0.82610 & 0.82259 & 0.82469 \\
        \end{tabular}
    \end{ruledtabular}
\end{table*}

All results presented here correspond to the exponential regulator \eqref{eq:regulator-exp}. Nonetheless, also the regulator family given in Eq.~\eqref{eq:regulator-wetterich} was considered, finding quantitative and qualitative agreement with the results just presented. For completeness, we present in Table \ref{tab:critExpWett} analogous results to those presented in Tab.~\ref{tab:critExpExp}. It pops out from the data and figures that the restoration of conformal symmetry is linked to a lower sensitivity to the regulator profile.

\begin{table*}[t]
    \caption{Results obtained using the Wetterich regulator given in Eq.~\eqref{eq:regulator-wetterich} for critical exponents using different criteria for fixing the regulator parameter $\alpha$. The error bar for PMS data is calculated just using the difference between the raw results for this regulator at order $O(\partial^4)$ and at order $O(\partial^2)$ divided by 4 (see \cite{DePolsi2020a}).}\label{tab:critExpWett}
    \begin{ruledtabular}
        \begin{tabular}{lclccccc}
         \textbf{} & \textbf{$\partial^4$} & & \textbf{$\mathcal{C}^{(1,1)}_{1}$} & \textbf{$\mathcal{C}^{(1,1)}_{2}$} & \textbf{$\mathcal{C}^{(2,1)}_{1}$} & \textbf{$\mathcal{C}^{(2,1)}_{2}$} & \textbf{$\mathcal{C}^{(3,1)}$} \\
         \colrule
         \textbf{$\nu$} & 0.63028(51) & \textbf{$\nu$, $\alpha_{PMC,0}$} & 0.63028 & 0.63028 & 0.63034 & 0.63139 & 0.63034 \\
         & &  \textbf{$\nu$, $\alpha_{PMC,\|\cdot\|}$} &  0.63028 & 0.63034 & 0.63040 & 0.63028 & 0.63028 \\
         \hline
         \textbf{$\omega$} & 0.8265(54) & \textbf{$\omega$, $\alpha_{PMC,0}$}  & 0.82380 & 0.82646 & 0.82335 & 0.82335 & 0.82380 \\
         & &  \textbf{$\omega$, $\alpha_{PMC,\|\cdot\|}$} & 0.82380 & 0.82646 & 0.82636 & 0.82380 & 0.82502 \\
        \end{tabular}
    \end{ruledtabular}
\end{table*}

\section{Conclusions}\label{Sec:Concl}

We have studied the violation of Ward identities when the DE is implemented at next-to-next-to-leading order, or $O(\partial^4)$, in the presence of a source for composite operators. By doing so, we were able to address two aspects: i) the study of four new conformal constraints that were not previously accessible in past implementations; and ii) the behavior of one particular constraint at successive orders. Both of these aspects are of much importance since, in general, when computing critical quantities, their exact values are typically not known and consequently one could doubt an approximation scheme which lacks  a rigorous bedrock. However, conformal symmetry restrictions should be in fact satisfied and, therefore, accessing the information stemming from them allows for decisive evidence with regard to convergence and accuracy of the approximation scheme. Moreover, these results imply an irrefutable criterion regarding the proper fixing of the scheme or regulator profile. This argument is not restricted to the DE approach implemented in this work, but applies to many others as well.

With regard to the first aspect, our investigation into the first set of constraints reveals that symmetry restrictions are most effectively satisfied when regulator profiles align with the PMS, yielding the same qualitative picture as previous studies. This observation reinforces the idea of equivalence between PMS and PMC criteria. This has the very important consequence of conferring on PMS a more sound base for implementation, given the fact that it is in practice a quicker and simpler implementation with a scope that goes beyond criticality. Moreover, experience shows that implementing PMS is crucial for the convergence of the DE.

The second aspect sheds light on a related issue, the behavior of the DE with increasing order. Indeed,  by tracking a specific constraint across successive orders of the DE, we were able to demonstrate two facts. First, these results provide compelling evidence that the DE is indeed converging, as this constraint is improving with increasing order of the approximation. Second, the violation of the conformal constraint at order $O(\partial^4)$ is reduced by a factor of order $1/4$ relative to the violation at order $O(\partial^2)$ in the vicinity of $\alpha_{\mathrm{PMC}}$. This result strongly supports the interpretation of the DE as a controlled expansion governed by a small parameter of order $1/4$ related to the mass-like behavior of the regulated theory.

In summary, these results provide strong evidence that the derivative expansion converges toward conformal consistency as higher-order operators are included. They also highlight a deep connection between regulator optimization and symmetry restoration at criticality. Moving forward, we intend to extend this analysis to $O(N)$ theories and explore even higher orders within the expansion to further refine these critical benchmarks.

\section*{Acknowledgments}

G.D.P. acknowledges support from the Programa de Desarrollo de las Ciencias Básicas (PEDECIBA) and from the grant of number FCE-3-2024-1-180709 of the Agencia Nacional de Investigación e Innovación (Uruguay).

\bibliographystyle{unsrt}
\bibliography{confCompDE4}

\end{document}